\documentclass[superscriptaddress,showpacs,twocolumn,preprintnumbers]{revtex4}

\usepackage{amsmath}

\usepackage{float}
\usepackage{graphicx}	
\usepackage{hyperref}
\usepackage{mathtools}
\usepackage{cancel}
\usepackage{multirow}
\usepackage{rotating}
\usepackage{physics}

\begin{document}

\title{Natural Understanding of Sterile Neutrino by Relativistic Equation}

\author{Keiichi Kimura}
\email{kimukei@eken.phys.nagoya-u.ac.jp}
\affiliation{Department of Physics, Nagoya University, Nagoya, 464-8602,
Japan}

\author{Akira Takamura}
\email{takamura@eken.phys.nagoya-u.ac.jp}
\affiliation{Department of Physics, Nagoya University, Nagoya, 464-8602,
Japan}
\affiliation{Department of Mathematics,
Toyota National College of Technology, Eisei-cho 2-1, Toyota-shi, 471-8525, Japan}

\date{\today}
\begin{abstract}
We derive the neutrino oscillation probabilities including sterile neutrinos by using the Dirac equation. 
If neutrinos have both the Dirac and the Majorana mass terms, 
left-handed neutrino $\nu_L$ and right-handed anti-neutrino $\nu_R^c$ enter the same multiplet 
and can transfer each other. 
Sterile neutrinos have been introduced by hand as the fourth generation neutrino 
in many papers, 
however, we can understand sterile neutrinos naturally in the framework of three generations. 
We also point out that the oscillations into sterile neutrinos strongly suggest the existence of 
both the Dirac and the Majorana mass terms, and for neutrinos to be the Majorana particles. 
\end{abstract} 

\maketitle

\section{Introduction}
The idea of neutrino-antineutrino oscillations proposed by Pontecorvo in 1957 \cite{Pontecorvo}. After the discovery of muon neutrino, Maki, Nakagawa and Sakata \cite{MNS} proposed the oscillations between neutrinos with different flavors in 1962, and the oscillations have been confirmed in the Super-Kamiokande atmospheric neutrino experiment in 1998 \cite{1998SK}.
Over the next 20 years, the understanding of the neutrino mass squared differences and the mixing angles
has greatly proceeded by the various experiments about solar neutrino \cite{SK, SNO, SK2},
long-baseline neutrino \cite{T2K, MINOS} and reactor neutrino \cite{KamLAND, DayaBay, RENO, DoubleChooz}
and we are getting the clue of the leptonic Dirac CP phase at present \cite{Dirac CP, NOvA}.
In order to estimate the value of the Dirac CP phase as precisely as possible,
the exact formulation of the oscillation probabilities including matter effect
has been developed \cite{Zaglauer, Ohlsson, KTY, Yokomakura0207, Yasuda}.

On the other hand, some anomalies have been reported in the LSND experiment \cite{LSND},
the MiniBooNE experiment \cite{MiniBooNE},
the reactor experiments \cite{reactor-anomaly}, the Galium experiments \cite{Galium-anomaly}
and recently in another reactor experiments \cite{NEOS, DANSS, NEUTRINO-4}.
These phenomena cannot be explained in the usual three-generation framework and may suggest the existence of
sterile neutrinos.
However, we have not reached a clear conclusion yet because the negative results for the existence of sterile neutrinos
are also obtained in some experiments
\cite{solar-limit,MINOS-limit,MINOS+,NOvA-limit,T2K-limit,Daya Bay,PROSPECT,STEREO,SK-limit,IceCube,ANTARES}.
At present, several experiments are planned or start data-taking in order to explore the truth
\cite{JSNS2, SBN, CCM}. It is expected to clarify the existence of sterile neutrinos
and find new physics beyond the Standard Model.

We have extended the formulation of neutrino oscillations to relativistic ones
by using the Dirac equation in our previous papers \cite{KT1,KT2,KT3}.
The existence of neutrino mass has established by various experiments
but it is not clarified whether the mass is originated from the Dirac mass or the Majorana mass or both.
We derived the neutrino oscillation probabilities in the case with only the Dirac mass \cite{KT1},
with only the Majorana mass \cite{KT2} in two generations and in three generations or more \cite{KT3}.
If neutrinos have the Dirac mass term, $\nu_L \leftrightarrow \nu_R$ and $\nu_L^c \leftrightarrow \nu_R^c$ oscillations
occur by extending to the relativistic formulation.
In the sense that $\nu_R$ and $\nu_R^c$ do not have weak interactions in the framework of the Standard Model,
they are nothing but sterile neutrinos.
Namely, active neutrinos can oscillate to sterile neutrinos but the probabilities strongly suppressed
by the factor $m^2/E^2$ because the oscillations accompany the chirality-flip.
On the other hand, in the case with only the Majorana mass, there exist only $\nu_L$ and $\nu_L^c$
in the theory. Namely, sterile neutrinos do not appear.

In this paper, we derive the oscillation probabilities in the case with both the Dirac and the Majorana mass terms.
The existence of two kinds of mass terms open new oscillation channels $\nu_L \leftrightarrow \nu_R^c$ and
$\nu_L^c \leftrightarrow \nu_R$.
These new channels are not suppressed by the factor $m^2/E^2$ such as
$\nu_L \leftrightarrow \nu_R$ and $\nu_L^c \leftrightarrow \nu_R^c$ oscillations
because of the oscillations without chirality-flip.
Therefore, the oscillations to sterile neutrinos suggested in some experiments can be naturally explained
in this scenario.
In other words, the existence of the oscillations to sterile neutrinos provides the strong suggestion
that neutrino is a Majorana particle and has both the Dirac and the Majorana mass terms at the same time.
In previous papers, the sterile neutrino was often introduced as a fourth-generation neutrino
independent of three light active neutrinos.
However, it is considered to be natural that there are also three sterile neutrinos corresponding to
three active neutrinos and there are some relations between them.
More concretely speaking, it should be considered that the light ones of the right-handed anti-neutrinos $\nu_R^c$
included in the three-generation framework behave like fourth-generation neutrinos.

This paper is organized as follows.
In section 2, we describe the notation used in this paper.
In section 3, we consider one generation neutrino with both the Dirac and the Majorana mass terms
and derive the oscillation probabilities to explain the basic structure of our formulation.
In section 4, we extend our formulation to two or more generation neutrino.
In section 5, we summarize our results.

\section{Notation}
In this section, we write down the notation used in this paper.
We mainly use the chiral representation
because neutrinos are measured through weak interactions.
In chiral representation, the gamma matrices with $4\times 4$ form are given by
\begin{eqnarray}
\gamma^0=\left(\begin{array}{cc}0 & 1 \\ 1 & 0\end{array}\right), \,
\gamma^i=\left(\begin{array}{cc}0 & -\sigma_i \\ \sigma_i & 0\end{array}\right), \,
\gamma_5=\left(\begin{array}{cc}1 & 0 \\ 0 & -1\end{array}\right), \label{gamma-mat}
\end{eqnarray}
where $2\times 2$ $\sigma_i$ matrices are defined by
\begin{eqnarray}
\sigma_1=\left(\begin{array}{cc}0 & 1 \\ 1 & 0\end{array}\right), \,
\sigma_2=\left(\begin{array}{cc}0 & -i \\ i & 0\end{array}\right), \,
\sigma_3=\left(\begin{array}{cc}1 & 0 \\ 0 & -1\end{array}\right).
\end{eqnarray}
We also define 4-component spinors $\psi$, $\psi_L$ and $\psi_R$ as
\begin{eqnarray}
&&\hspace{-0.5cm} \psi=\left(\begin{array}{c}\xi \\ \eta \end{array}\right), \\
&&\hspace{-0.5cm} \psi_L=\frac{1-\gamma_5}{2}\psi=\left(\begin{array}{c}0 \\ \eta\end{array}\right), \,
\psi_R=\frac{1+\gamma_5}{2}\psi=\left(\begin{array}{c}\xi \\ 0 \end{array}\right), \label{psi-def}
\end{eqnarray}
and 2-component spinors $\xi$ and $\eta$ as
\begin{eqnarray}
\xi=\left(\begin{array}{c}\nu_R^{\prime} \\ \nu_R \end{array}\right), \qquad
\eta=\left(\begin{array}{c}\nu_L^{\prime} \\ \nu_L \end{array}\right).
\end{eqnarray}
Furthermore,
we use the subscript $\alpha$ and $\beta$ for flavor, $L$ and $R$ for chirality,
the number $j$ and $k$ for generation and superscript $\pm$ for energy.
Because of negligible neutrino mass, mass eigenstate has been often identified with energy eigenstate
in the previous papers.
But in the future, we should distinguish these two kinds of eigenstates for the finite neutrino mass.
More concretely, we use the following eigenstates;
\begin{eqnarray}
&&\hspace{-0.5cm}
{\rm chirality\mathchar`-flavor \,\, eigenstates}: \quad \nu_{\alpha L}, \nu_{\alpha R}, \nu_{\beta L}, \nu_{\beta R}, \\
&&\hspace{-0.5cm}{\rm chirality\mathchar`-mass \,\, eigenstates}: \quad\,\, \nu_{jL}, \nu_{jR}, \nu_{kL}, \nu_{kR}, \\
&&\hspace{-0.5cm}{\rm energy\mathchar`-helicity \,\, eigenstates}: \quad \nu_j^+, \nu_j^-, \nu_k^+, \nu_k^-.
\end{eqnarray}
It is noted that chirality-mass eigenstates are not exactly the eigenstates of the Hamiltonian.
We use the term, eigenstates, in the sense that the mass submatrix in the Hamiltonian is diagonalized.
Judging from common sense, one may think it strange that the chirality and the mass live in the same eigenstate.
Details will be explained in the subsequent section.

We also difine the spinor for anti-neutrino as charge conjugation of neutrino $\psi^c=i\gamma^2 \psi^*$.
The charge conjugations for left-handed and right-handed neutrinos are defined by
\begin{eqnarray}
&&\hspace{-0.5cm}\psi_L^c\equiv(\psi_L)^c\equiv
\left(\begin{array}{c}\nu_L^c \\ \nu_L^{c\prime} \\ 0 \\ 0 \end{array}\right)
\equiv i\gamma^2 \psi_L^*=i\gamma^2 \frac{1-\gamma_5}{2}\psi^* \nonumber \\
&&\hspace{-0.5cm}=\frac{1+\gamma_5}{2}(i\gamma^2 \psi^*)
=(\psi^c)_R=\left(\!\!\begin{array}{c}i\sigma_2 \eta^* \\ 0 \end{array}\!\!\right)
=\left(\!\!\begin{array}{c}\nu_L^* \\ -\nu_L^{*\prime} \\ 0 \\ 0 \end{array}\!\!\right), \label{nuc} \\
&&\hspace{-0.5cm}\psi_R^c\equiv(\psi_R)^c\equiv\left(\begin{array}{c}0 \\ 0 \\ \nu_R^c \\ \nu_R^{c\prime} \end{array}\right)
\equiv i\gamma^2 \psi_R^*=i\gamma^2 \frac{1+\gamma_5}{2}\psi^* \nonumber \\
&&\hspace{-0.5cm}=\frac{1-\gamma_5}{2}(i\gamma^2 \psi^*)
=(\psi^c)_L=\left(\!\!\!\begin{array}{c}0 \\ -i\sigma_2 \xi^* \end{array}\!\!\!\right)
=\left(\!\!\begin{array}{c}0 \\ 0 \\ -\nu_R^* \\ \nu_R^{*\prime} \end{array}\!\!\right).
\end{eqnarray}
It is noted that the chirality is flipped by taking the charge conjugation.

\section{Oscillation Probabilities in one generation}

In this section, let us calculate the oscillation probabilities in one generation by using the Dirac equation,
which is a relativistic equation. In particular, we consider the case with both the Dirac and the Majorana mass terms.

We can introduce the Majorana mass for both left-handed and right-handed neutrinos theoretically.
The triplet Higgs is needed to introduce the Majorana mass for left-handed neutrinos
as they have weak interactions.
However, the triplet Higgs has not been found yet at present.
So, it seems that the Majorana mass term for left-handed neutrinos does not exist experimentally.
On the other hand, the Majorana mass for right-handed neutrinos does not subject to theoretical constraints
as they do not have weak interactions.

Here, let us consider the general case that both the left-handed and right-handed neutrinos have
the Majorana mass terms. The lagrangian, in this case, is given by
\begin{eqnarray}
L&=&\frac{1}{2}i\overline{\psi_L}\gamma^{\mu}\partial_{\mu}\psi_L
+\frac{1}{2}i\overline{\psi_R}\gamma^{\mu}\partial_{\mu}\psi_R \nonumber \\
&&+\frac{1}{2}i\overline{\psi_L^{c}}\gamma^{\mu}\partial_{\mu}\psi_L^{c}
+\frac{1}{2}i\overline{\psi_R^{c}}\gamma^{\mu}\partial_{\mu}\psi_R^{c} \nonumber \\
&&-\overline{\psi_L}m^*\psi_R-\overline{\psi_R}m\psi_L
-\overline{\psi_L^c}m^*\psi_R^c-\overline{\psi_R^c}m\psi_L^c \nonumber \\
&&-\frac{1}{2}\overline{\psi_L}M_L\psi_L^c
-\frac{1}{2}\overline{\psi_L^c}M_L^*\psi_L \nonumber \\
&&-\frac{1}{2}\overline{\psi_R}M_R\psi_R^c
-\frac{1}{2}\overline{\psi_R^c}M_R^*\psi_R,
\end{eqnarray}
where $m$ is the Dirac mass and $M_L$, $M_R$ are the Majorana masses for left-handed and right-handed neutrinos.
From the Eular-Lagrange equation for $\overline{\psi_R}$
\begin{eqnarray}
\frac{\partial L}{\partial \overline{\psi_R}}
-\partial_\mu \left(\frac{\partial L}{\partial(\partial_\mu \overline{\psi_R})}\right)=0,
\end{eqnarray}
the Dirac equation
\begin{eqnarray}
i\gamma^\mu \partial_\mu \psi_R-m \psi_L
-M_R\psi_R^c=0
\end{eqnarray}
is obtained.
Multiplying $\gamma_0$ from the left, the above equation becomes
\begin{eqnarray}
i\partial_0 \psi_R+i\gamma^0\gamma^i\partial_i \psi_R
-m \gamma^0\psi_L-M_R\gamma^0 \psi_R^c=0. \label{dirac-eqR}
\end{eqnarray}
In the same way, we obtain the quation for $\psi_L$,
\begin{eqnarray}
i\partial_0 \psi_L+i\gamma^0\gamma^i\partial_i \psi_L
-m \gamma^0\psi_R-M_L\gamma^0 \psi_L^c=0. \label{dirac-eqL}
\end{eqnarray}
Here, we set the equal momentum assumption and factor out the dependence of the
distance as
\begin{eqnarray}
\psi(x,t)=e^{i\vec{p}\cdot \vec{x}}\psi(t),
\end{eqnarray}
and we choose $\overrightarrow{p}=(0,0,p)$.
If we write these equations (\ref{dirac-eqR}) and (\ref{dirac-eqL}) together in a matrix form,
the time evolution of the chirality-flavor eigenstates is
represented as
\begin{widetext}
\begin{eqnarray}
i\frac{d}{dt}
\left(\begin{array}{l}
\nu_{L}^{c} \\ \nu_{L}^{c\prime} \\ \nu_{L}^{\prime} \\ \nu_{L} \\
\nu_{R}^{\prime} \\ \nu_{R} \\ \nu_{R}^{c} \\ \nu_{R}^{c\prime}
\end{array}\right)
=\left(\begin{array}{cccc|cccc}
-p & 0 & M_L & 0 & 0 & 0 & m & 0 \\
0 & p & 0 & M_L & 0 & 0 & 0 & m \\
M_L^* & 0 & -p & 0 & m^* & 0 & 0 & 0 \\
0 & M_L^* & 0 & p & 0 & m^* & 0 & 0 \\
\hline
0 & 0 & m & 0 & p & 0 & M_R & 0 \\
0 & 0 & 0 & m & 0 & -p & 0 & M_R \\
m^* & 0 & 0 & 0 & M_R^* & 0 & p & 0 \\
0 & m^* & 0 & 0 & 0 & M_R^* & 0 & -p
\end{array}\right)
\left(\begin{array}{l}
\nu_{L}^{c} \\ \nu_{L}^{c\prime} \\ \nu_{L}^{\prime} \\ \nu_{L} \\
\nu_{R}^{\prime} \\ \nu_{R} \\ \nu_{R}^{c} \\ \nu_{R}^{c\prime}
\end{array}\right). \label{dirac-eq1}
\end{eqnarray}
Exchanging some rows and some columns in the above matrix, we can separate to two parts completely as follows,
\begin{eqnarray}
i\frac{d}{dt}\left(\begin{array}{l}
\nu_L^{c\prime} \\ \nu_R^{\prime} \\ \nu_L^{\prime} \\ \nu_R^{c\prime} \\
\nu_L^{c} \\ \nu_R \\ \nu_L \\ \nu_R^{c}
\end{array}\right)
=\left(\begin{array}{cccc|cccc}
p & 0 & M_L & m & 0 & 0 & 0 & 0 \\
0 & p & m & M_R & 0 & 0 & 0 & 0 \\
M_L^* & m^* & -p & 0 & 0 & 0 & 0 & 0 \\
m^* & M_R^* & 0 & -p & 0 & 0 & 0 & 0 \\
\hline
0 & 0 & 0 & 0 & -p & 0 & M_L & m \\
0 & 0 & 0 & 0 & 0 & -p & m & M_R \\
0 & 0 & 0 & 0 & M_L^* & m^* & p & 0 \\
0 & 0 & 0 & 0 & m^* & M_R^* & 0 & p
\end{array}\right)\left(\begin{array}{l}
\nu_L^{c\prime} \\ \nu_R^{\prime} \\ \nu_L^{\prime} \\ \nu_R^{c\prime} \\
\nu_L^{c} \\ \nu_R \\ \nu_L \\ \nu_R^{c}
\end{array}\right). \label{dirac-eq2}
\end{eqnarray}
\end{widetext}
Taking out the lower-right part of this equation, we obtain
\begin{eqnarray}
i\frac{d}{dt}\left(\begin{array}{l}
\nu_L^{c} \\ \nu_R \\ \nu_L \\ \nu_R^{c}
\end{array}\right)
=\left(\begin{array}{cc|cc}
-p & 0 & M_L & m \\
0 & -p & m & M_R \\
\hline
M_L^* & m^* & p & 0 \\
m^* & M_R^* & 0 & p
\end{array}\right)\left(\begin{array}{l}
\nu_L^{c} \\ \nu_R \\ \nu_L \\ \nu_R^{c}
\end{array}\right). \label{H-one-gene}
\end{eqnarray}
This has exactly the same structure as two-generation neutrino case with only the Majorana mass.
Although negative mass eigenvalue appears by diagonalizing the mass matrix in the seesaw mechanism,
it can be changed to positive by using the phase redefinition of a neutrino field.

Here, let us some comments on the above equation.
In the case that the Majorana masses $M_L$ and $M_R$ are equal to zero,
only $\nu_L \leftrightarrow \nu_R$ and $\nu_L^c \leftrightarrow \nu_R^c$ oscillations can occur in addition to
usual $\nu_L \leftrightarrow \nu_L$ oscillations and it returns to the case with only the Dirac mass \cite{KT1}.
The lepton number also conserved in this case.
In the case that the Dirac mass $m$ is equal to zero,
only $\nu_L \leftrightarrow \nu_L^c$ and $\nu_R \leftrightarrow \nu_R^c$ oscillations can occur in addition to
$\nu_L \leftrightarrow \nu_L$ oscillations and it returns to the case with only the Majorana masses \cite{KT2}.
It is found that $\nu_L \leftrightarrow \nu_R^c$ and $\nu_R \leftrightarrow \nu_L^c$ oscillations are realized
only when there are both types of mass terms.
In the framework of the Standard Model, $\nu_R$ and $\nu_R^c$ do not interact with other particles
through weak interactions and interact only through gravity, so they are called sterile neutrinos.
Namely, if there exist both the Dirac and the Majorana mass terms, active neutrinos oscillate to sterile neutrinos,
and $\nu_R^c$ and $\nu_R$ play the role as if they were the fourth-generation neutrinos.
Conversely, if one can confirm the oscillations to sterile neutrinos, it becomes a strong suggestion for neutrinos
to have both the Dirac and the Majorana masses and to be a Majorana particle.

Next, let us calculate the oscillation probabilities by diagonalizing the Hamiltonian in (\ref{H-one-gene}).
We diagonalize the mass matrix in the Hamiltonian first and then the whole.
The first step is performed as follows,
\begin{widetext}
\begin{eqnarray}
\left(\begin{array}{cc|cc}
U_{\alpha 1} & U_{\bar{\alpha}1} & 0 & 0 \\
U_{\alpha 2} & U_{\bar{\alpha}2} & 0 & 0 \\
\hline
0 & 0 & U_{\alpha 1}^* & U_{\bar{\alpha}1}^* \\
0 & 0 & U_{\alpha 2}^* & U_{\bar{\alpha}2}^*
\end{array}\right)
\left(\begin{array}{cc|cc}
-p & 0 & M_L & m \\
0 & -p & m & M_R \\
\hline
M_L^* & m^* & p & 0 \\
m^* & M_R^* & 0 & p
\end{array}\right)
\left(\begin{array}{cc|cc}
U_{\alpha 1}^* & U_{\alpha 2}^* & 0 & 0 \\
U_{\bar{\alpha}1}^* & U_{\bar{\alpha}2}^* & 0 & 0 \\
\hline
0 & 0 & U_{\alpha 1} & U_{\alpha 2} \\
0 & 0 & U_{\bar{\alpha}1} & U_{\bar{\alpha}2}
\end{array}\right)
=\left(\begin{array}{cc|cc}
-p & 0 & m_1 & 0 \\
0 & -p & 0 & m_2 \\
\hline
m_1 & 0 & p & 0 \\
0 & m_2 & 0 & p
\end{array}\right),
\label{diag1}
\end{eqnarray}
where $U$ is a unitary matrix and $m_1$ and $m_2$ are mass eigenvalues.
The mass matrix $M$ is diagonalized in the form $U^T MU$ as it is complex symmetric.
$U$ multiplied a phase matrix can also diagonalize the mass matrix in the same way.
So, we can choose $m_1$ and $m_2$ as positive by using the freedom of this phase redefinition.
The chirality-flavor eigenstates are related to the chirality-mass eigenstates as
\begin{eqnarray}
\left(\begin{array}{l}
\nu_{L}^{c} \\ \nu_{R} \\ \nu_{L} \\ \nu_{R}^{c}
\end{array}\right)
=\left(\begin{array}{cc|cc}
U_{\alpha 1}^* & U_{\alpha 2}^* & 0 & 0 \\
U_{\bar{\alpha}1}^* & U_{\bar{\alpha}2}^* & 0 & 0 \\
\hline
0 & 0 & U_{\alpha 1} & U_{\alpha 2} \\
0 & 0 & U_{\bar{\alpha}1} & U_{\bar{\alpha}2}
\end{array}\right)\left(\begin{array}{l}
\nu_{1L}^{c} \\ \nu_{2R} \\ \nu_{1L} \\ \nu_{2R}^{c}
\end{array}\right).\label{CF-CM}
\end{eqnarray}
Substituting (\ref{CF-CM}) into (\ref{H-one-gene}), we obtain the evolution equation for mass eigenstates,
\begin{eqnarray}
i\frac{d}{dt}\left(\begin{array}{l}
\nu_{1L}^{c} \\ \nu_{2R} \\ \nu_{1L} \\ \nu_{2R}^{c}
\end{array}\right)
=\left(\begin{array}{cccc}
-p & 0 & m_1 & 0 \\
0 & -p & 0 & m_2 \\
m_1 & 0 & p & 0 \\
0 & m_2 & 0 & p
\end{array}\right)\left(\begin{array}{l}
\nu_{1L}^{c} \\ \nu_{2R} \\ \nu_{1L} \\ \nu_{2R}^{c}
\end{array}\right). \label{te-CM}
\end{eqnarray}
Furthermore, we would like to rewrite it to the equation for energy-helicity eigenstates
in order to diagonalize the Hamiltonian in (\ref{te-CM}) completely.
Exchaging some rows and some columns, the above equation becomes
\begin{eqnarray}
i\frac{d}{dt}\left(\begin{array}{l}
\nu_{1L}^{c} \\ \nu_{1L} \\ \nu_{2R} \\ \nu_{2R}^{c}
\end{array}\right)
=\left(\begin{array}{cc|cc}
-p & m_1 & 0 & 0 \\
m_1 & p & 0 & 0 \\\hline
0 & 0 & -p & m_2 \\
0 & 0 & m_2 & p
\end{array}\right)\left(\begin{array}{l}
\nu_{1L}^{c} \\ \nu_{1L} \\ \nu_{2R} \\ \nu_{2R}^{c}
\end{array}\right).
\end{eqnarray}
The chirality-mass eigenstates are given by the linear combination of the energy-helicity eigenstates as
\begin{eqnarray}
\left(\begin{array}{c}
\nu_{1L}^{c} \\ \nu_{1L} \\ \nu_{2R} \\ \nu_{2R}^{c}
\end{array}\right)
=\left(\begin{array}{cc|cc}
\sqrt{\frac{E_1+p}{2E_1}} & \sqrt{\frac{E_1-p}{2E_1}} & 0 & 0 \\
-\sqrt{\frac{E_1-p}{2E_1}} & \sqrt{\frac{E_1+p}{2E_1}} & 0 & 0 \\
\hline
0 & 0 & \sqrt{\frac{E_2+p}{2E_2}} & \sqrt{\frac{E_2-p}{2E_2}} \\
0 & 0 & -\sqrt{\frac{E_2-p}{2E_2}} & \sqrt{\frac{E_2+p}{2E_2}}
\end{array}\right)\left(\begin{array}{l}
\nu_{1}^{c-} \\ \nu_{1}^+ \\ \nu_{2}^{-} \\ \nu_{2}^{c+}
\end{array}\right), \label{CM-EH}
\end{eqnarray}
and the Hamitonian in the evolution equation for the energy-helicity eigenstates are completely diagonalized as
\begin{eqnarray}
i\frac{d}{dt}\left(\begin{array}{l}
\nu_{1}^{c-} \\ \nu_{1}^+ \\ \nu_{2}^{-} \\ \nu_{2}^{c+}
\end{array}\right)
=\left(\begin{array}{cccc}
-E_1 & 0 & 0 & 0 \\
0 & E_1 & 0 & 0 \\
0 & 0 & -E_2 & 0 \\
0 & 0 & 0 & E_2
\end{array}\right)\left(\begin{array}{l}
\nu_{1}^{c-} \\ \nu_{1}^+ \\ \nu_{2}^{-} \\ \nu_{2}^{c+}
\end{array}\right),
\end{eqnarray}
where the eigenvalues are given by
\begin{eqnarray}
E_j&=&\sqrt{p^2+m_j^2} \quad (j=1,2).
\end{eqnarray}
Connecting equations (\ref{CF-CM}) and (\ref{CM-EH}),
the chirality-flavor eigenstates are expressed by the energy-helicity eigenstates as
\begin{eqnarray}
\left(\begin{array}{l}
\nu_{L}^{c} \\ \nu_{R} \\ \nu_{L} \\ \nu_{R}^{c}
\end{array}\right)
&=&
\left(\begin{array}{cc|cc}
U_{\alpha 1}^{*} & U_{\alpha 2}^{*} & 0 & 0 \\
U_{\bar{\alpha}1}^{*} & U_{\bar{\alpha}2}^{*} & 0 & 0 \\
\hline
0 & 0 & U_{\alpha 1} & U_{\alpha 2} \\
0 & 0 & U_{\bar{\alpha}1} & U_{\bar{\alpha}2}
\end{array}\right)
\left(\begin{array}{cccc}
1 & 0 & 0 & 0 \\
0 & 0 & 1 & 0 \\
0 & 1 & 0 & 0 \\
0 & 0 & 0 & 1
\end{array}\right)
\left(\begin{array}{cc|cc}
\sqrt{\frac{E_1+p}{2E_1}} & \sqrt{\frac{E_1-p}{2E_1}} & 0 & 0 \\
-\sqrt{\frac{E_1-p}{2E_1}} & \sqrt{\frac{E_1+p}{2E_1}} & 0 & 0 \\
\hline
0 & 0 & \sqrt{\frac{E_2+p}{2E_2}} & \sqrt{\frac{E_2-p}{2E_2}} \\
0 & 0 & -\sqrt{\frac{E_2-p}{2E_2}} & \sqrt{\frac{E_2+p}{2E_2}}
\end{array}\right)\left(\begin{array}{l}
\nu_{1}^{c-} \\ \nu_{1}^+ \\ \nu_{2}^{-} \\ \nu_{2}^{c+}
\end{array}\right) \nonumber \\
&=&\left(\begin{array}{cc|cc}
\sqrt{\frac{E_1+p}{2E_1}}U_{\alpha 1}^* & \sqrt{\frac{E_1-p}{2E_1}}U_{\alpha 1}^*
& \sqrt{\frac{E_2+p}{2E_2}}U_{\alpha 2}^* & \sqrt{\frac{E_2-p}{2E_2}}U_{\alpha 2}^* \\
\sqrt{\frac{E_1+p}{2E_1}}U_{\bar{\alpha}1}^* & \sqrt{\frac{E_1-p}{2E_1}}U_{\bar{\alpha}1}^*
& \sqrt{\frac{E_2+p}{2E_2}}U_{\bar{\alpha}2}^* & \sqrt{\frac{E_2-p}{2E_2}}U_{\bar{\alpha}2}^* \\
\hline
-\sqrt{\frac{E_1-p}{2E_1}}U_{\alpha 1} & \sqrt{\frac{E_1+p}{2E_1}}U_{\alpha 1}
& -\sqrt{\frac{E_2-p}{2E_2}}U_{\alpha 2} & \sqrt{\frac{E_2+p}{2E_2}}U_{\alpha 2} \\
-\sqrt{\frac{E_1-p}{2E_1}}U_{\bar{\alpha}1} & \sqrt{\frac{E_1+p}{2E_1}}U_{\bar{\alpha}1}
& -\sqrt{\frac{E_2-p}{2E_2}}U_{\bar{\alpha}2} & \sqrt{\frac{E_2+p}{2E_2}}U_{\bar{\alpha}2}
\end{array}\right)
\left(\begin{array}{l}
\nu_{1}^{c-} \\ \nu_{1}^+ \\ \nu_{2}^{-} \\ \nu_{2}^{c+}
\end{array}\right).
\end{eqnarray}
Rewriting these relations for fields to those for one particle states, the chirality-mass eigenstates after the
time $t$ are given by
\begin{eqnarray}
&&\hspace{-1cm}|\nu_L^{c}(t)\rangle =\sqrt{\frac{E_1+p}{2E_1}}U_{\alpha 1}e^{iE_1t}|\nu_1^{c-}\rangle
+\sqrt{\frac{E_1-p}{2E_1}}U_{\alpha 1}e^{-iE_1t}|\nu_1^{+}\rangle
+\sqrt{\frac{E_2+p}{2E_2}}U_{\alpha 2}e^{iE_2}|\nu_2^{-}\rangle
+\sqrt{\frac{E_2-p}{2E_2}}U_{\alpha 2}e^{-iE_2t}|\nu_2^{c+}\rangle, \\
&&\hspace{-1cm}|\nu_R(t)\rangle =\sqrt{\frac{E_1+p}{2E_1}}U_{\bar{\alpha}1}e^{iE_1t}|\nu_1^{c-}\rangle
+\sqrt{\frac{E_1-p}{2E_1}}U_{\bar{\alpha}1}e^{-iE_1t}|\nu_1^{+}\rangle
+\sqrt{\frac{E_2+p}{2E_2}}U_{\bar{\alpha}2}e^{iE_2t}|\nu_2^{-}\rangle
+\sqrt{\frac{E_2-p}{2E_2}}U_{\bar{\alpha}2}e^{-iE_2t}|\nu_2^{c+}\rangle, \\
&&\hspace{-1cm}|\nu_L(t)\rangle =-\sqrt{\frac{E_1-p}{2E_1}}U_{\alpha 1}^*e^{iE_1t}|\nu_1^{c-}\rangle
\!+\!\sqrt{\frac{E_1+p}{2E_1}}U_{\alpha 1}^*e^{-iE_1t}|\nu_1^{+}\rangle
\!-\!\sqrt{\frac{E_2-p}{2E_2}}U_{\alpha 2}^*e^{iE_2t}|\nu_2^{-}\rangle
\!+\!\sqrt{\frac{E_2+p}{2E_2}}U_{\alpha 2}^*e^{-iE_2t} |\nu_2^{c+}\rangle, \\
&&\hspace{-1cm}|\nu_R^{c}(t)\rangle =-\sqrt{\frac{E_1-p}{2E_1}}U_{\bar{\alpha}1}^*e^{iE_1t}|\nu_1^{c-}\rangle
\!+\!\sqrt{\frac{E_1+p}{2E_1}}U_{\bar{\alpha}1}^*e^{-iE_1t}|\nu_1^{+}\rangle
\!-\!\sqrt{\frac{E_2-p}{2E_2}}U_{\bar{\alpha}2}^*e^{iE_2t}|\nu_2^{-}\rangle
\!+\!\sqrt{\frac{E_2+p}{2E_2}}U_{\bar{\alpha}2}^*e^{-iE_2t}|\nu_2^{c+}\rangle ,
\end{eqnarray}
and the conjugate states are also given by
\begin{eqnarray}
\langle \nu_L^{c}| &=&\sqrt{\frac{E_1+p}{2E_1}}U_{\alpha 1}^*\langle \nu_1^{c-}|
+\sqrt{\frac{E_1-p}{2E_1}}U_{\alpha 1}^*\langle \nu_1^+|+\sqrt{\frac{E_2+p}{2E_2}}U_{\alpha 2}^*\langle \nu_2^{-}|
+\sqrt{\frac{E_2-p}{2E_2}}U_{\alpha 2}^*\langle \nu_2^{c+}|, \\
\langle \nu_R| &=&\sqrt{\frac{E_1+p}{2E_1}}U_{\bar{\alpha}1}^*\langle \nu_1^{c-}|
+\sqrt{\frac{E_1-p}{2E_1}}U_{\bar{\alpha}1}^*\langle \nu_1^+| +\sqrt{\frac{E_2+p}{2E_2}}U_{\bar{\alpha}2}^*\langle \nu_2^{-}|
+\sqrt{\frac{E_2-p}{2E_2}}U_{\bar{\alpha}2}^*\langle \nu_2^{c+}|, \\
\langle \nu_L| &=&-\sqrt{\frac{E_1-p}{2E_1}}U_{\alpha 1}\langle \nu_1^{c-}|
+\sqrt{\frac{E_1+p}{2E_1}}U_{\alpha 1}\langle \nu_1^+|-\sqrt{\frac{E_2-p}{2E_2}}U_{\alpha 2}\langle \nu_2^{-}|
+\sqrt{\frac{E_2+p}{2E_2}}U_{\alpha 2}\langle \nu_2^{c+}|, \\
\langle \nu_R^{c}| &=&-\sqrt{\frac{E_1-p}{2E_1}}U_{\bar{\alpha}1}\langle \nu_1^{c-}|
+\sqrt{\frac{E_1+p}{2E_1}}U_{\bar{\alpha}1}\langle \nu_1^+| -\sqrt{\frac{E_2-p}{2E_2}}U_{\bar{\alpha}2}\langle \nu_2^{-}|
+\sqrt{\frac{E_2+p}{2E_2}}U_{\bar{\alpha}2}\langle \nu_2^{c+}|.
\end{eqnarray}
Then, the oscillation amplitudes after the time $t$ are given by
\begin{eqnarray}
A(\nu_L\to\nu_L)&=&|U_{\alpha 1}|^2\left(\frac{E_1+p}{2E_1}e^{-iE_1t}+\frac{E_1-p}{2E_1}e^{iE_1t}\right)
+|U_{\alpha 2}|^2\left(\frac{E_2+p}{2E_2}e^{-iE_2t}+\frac{E_2-p}{2E_2}e^{iE_2t}\right) \nonumber \\
&=&|U_{\alpha 1}|^2\left\{\cos(E_1t)-i\frac{p}{E_1}\sin (E_1t)\right\}
+|U_{\alpha 2}|^2\left\{\cos(E_2t)-i\frac{p}{E_2}\sin (E_2t)\right\}, \\
A(\nu_L\to\nu_R^{c})&=&U_{\alpha 1}^*U_{\bar{\alpha}1}\left(\frac{E_1+p}{2E_1}e^{-iE_1t}+\frac{E_1-p}{2E_1}e^{iE_1t}\right)
+U_{\alpha 2}^*U_{\bar{\alpha}2}\left(\frac{E_2+p}{2E_2}e^{-iE_2t}+\frac{E_2-p}{2E_2}e^{iE_2t}\right) \nonumber \\
&=&U_{\alpha 1}^*U_{\bar{\alpha}1}\left\{\cos(E_1t)-i\frac{p}{E_1}\sin (E_1t)\right\}
+U_{\alpha 2}^*U_{\bar{\alpha}2}\left\{\cos(E_2t)-i\frac{p}{E_2}\sin (E_2t)\right\}, \\
A(\nu_L\to\nu_L^{c})&=&U_{\alpha 1}^{*2}\frac{m_1}{2E_1}(e^{-iE_1t}-e^{iE_1t})
+U_{\alpha 2}^{*2}\frac{m_2}{2E_2}(e^{-iE_2t}-e^{iE_2t})
=-iU_{\alpha 1}^{*2}\frac{m_1}{E_1}\sin (E_1t)
-iU_{\alpha 2}^{*2}\frac{m_2}{E_2}\sin (E_2t), \\
A(\nu_L\to\nu_R)&=&U_{\alpha 1}^*U_{\bar{\alpha}1}^*\frac{m_1}{2E_1}(e^{-iE_1t}-e^{iE_1t})
+U_{\alpha 2}^*U_{\bar{\alpha}2}^*\frac{m_2}{2E_2}(e^{-iE_2t}-e^{iE_2t}) \nonumber \\
&=&-iU_{\alpha 1}^*U_{\bar{\alpha}1}^*\frac{m_1}{E_1}\sin (E_1t)-iU_{\alpha 2}^*U_{\bar{\alpha}2}^*\frac{m_2}{E_2}\sin (E_2t).
\end{eqnarray}
The oscillation probabilities are calculated by squaring these amplitudes as
\begin{eqnarray}
P(\nu_L\to\nu_L)&=&\left\{|U_{\alpha 1}|^2\cos(E_1t)+|U_{\alpha 2}|^2\cos(E_2t)\right\}^2
+\left\{|U_{\alpha 1}|^2\frac{p}{E_1}\sin (E_1t)
+|U_{\alpha 2}|^2\frac{p}{E_2}\sin (E_2t)\right\}^2, \label{P1} \\
P(\nu_L\to\nu_R^{c})&=&\left|U_{\alpha 1}U_{\bar{\alpha}1}^*\left\{\cos(E_1t)-i\frac{p}{E_1}\sin (E_1t)\right\}\right|^2
+\left|U_{\alpha 2}U_{\bar{\alpha}2}^*\left\{\cos(E_2t)-i\frac{p}{E_2}\sin (E_2t)\right\}\right|^2 \nonumber \\
&&+2{\rm Re}\left[U_{\alpha 1}U_{\bar{\alpha}1}^*\left\{\cos(E_1t)-i\frac{p}{E_1}\sin (E_1t)\right\}
U_{\alpha 2}^*U_{\bar{\alpha}2}\left\{\cos(E_2t)+i\frac{p}{E_2}\sin (E_2t)\right\}\right] \nonumber \\
&=&\left|U_{\alpha 1}U_{\bar{\alpha}1}\right|^2\left\{\cos^2(E_1t)+\frac{p^2}{E_1^2}\sin^2 (E_1t)\right\}
+\left|U_{\alpha 2}U_{\bar{\alpha}2}\right|^2\left\{\cos^2(E_2t)+\frac{p^2}{E_2^2}\sin^2 (E_2t)\right\} \nonumber \\
&&+2{\rm Re}\left[U_{\alpha 1}U_{\bar{\alpha}1}^*\left\{\cos(E_1t)-i\frac{p}{E_1}\sin (E_1t)\right\}
U_{\alpha 2}^*U_{\bar{\alpha}2}\left\{\cos(E_2t)+i\frac{p}{E_2}\sin (E_2t)\right\}\right], \\
P(\nu_L\to\nu_L^{c})&=&\left|U_{\alpha 1}^2\frac{m_1}{E_1}\sin (E_1t)+U_{\alpha 2}^2\frac{m_2}{E_2}\sin (E_2t)\right|^2 \nonumber \\
&=&\left|U_{\alpha 1}^2\frac{m_1}{E_1}\sin (E_1t)\right|^2+\left|U_{\alpha 2}^2\frac{m_2}{E_2}\sin (E_2t)\right|^2
+2{\rm Re}\left(U_{\alpha 1}^2U_{\alpha 2}^{*2}\right)\frac{m_1m_2}{E_1E_2}\sin (E_1t)\sin (E_2t), \\
P(\nu_L\to\nu_R)&=&\left|U_{\alpha 1}U_{\bar{\alpha}1}\frac{m_1}{E_1}\sin (E_1t)+U_{\alpha 2}U_{\bar{\alpha}2}\frac{m_2}{E_2}\sin (E_2t)\right|^2 \nonumber \\
&&\hspace{-1cm}=\left|U_{\alpha 1}U_{\bar{\alpha}1}\frac{m_1}{E_1}\sin (E_1t)\right|^2+\left|U_{\alpha 2}U_{\bar{\alpha}2}\frac{m_2}{E_2}\sin (E_2t)\right|^2
+2{\rm Re}\left(U_{\alpha 1}U_{\bar{\alpha}1}U_{\alpha 2}^*U_{\bar{\alpha}2}^*\right)\frac{m_1m_2}{E_1E_2}\sin (E_1t)\sin (E_2t). \label{P4}
\end{eqnarray}
The $2\times 2$ unitary matrix has four parameters and can be parametrized as
\begin{eqnarray}
U&=&\left(\begin{array}{cc}
e^{i\rho_{\alpha}} & 0 \\
0 & e^{i\rho_{\bar{\alpha}}}
\end{array}\right)
\left(\begin{array}{cc}
\cos \theta & \sin \theta \\
-\sin \theta & \cos \theta
\end{array}\right)\left(\begin{array}{cc}
1 & 0 \\
0 & e^{i\phi}
\end{array}\right)
=
\left(\begin{array}{cc}
e^{i\rho_{\alpha}}\cos \theta & e^{i(\rho_{\alpha}+\phi)}\sin \theta \\
-e^{i\rho_{\bar{\alpha}}}\sin \theta & e^{i(\rho_{\bar{\alpha}}+\phi)}\cos \theta
\end{array}\right).
\end{eqnarray}
Substituting it into (\ref{P1})-(\ref{P4}), we obtain
\begin{eqnarray}
P(\nu_L\to\nu_L)&=&
1-4s^2c^2\sin^2 \frac{(E_2-E_1)t}{2} \label{L-L1}\\
&&-\left[c^4\cdot \frac{m_1^2}{E_1^2}\sin^2 (E_1t)
+s^4\cdot \frac{m_2^2}{E_2^2}\sin^2 (E_2t)+2s^2c^2\cdot \left(1-\frac{p^2}{E_1E_2}\right)
\sin (E_1t)\sin (E_2t)\right] \label{L-L2}, \\
P(\nu_L\to\nu_R^{c})&=&
4s^2c^2\sin^2 \frac{(E_2-E_1)t}{2} \label{L-Rc1}\\
&&-s^2c^2\left[\frac{m_1^2}{E_1^2}\sin^2 (E_1t)+\frac{m_2^2}{E_2^2}\sin^2 (E_2t)
-2\left(1-\frac{p^2}{E_1E_2}\right)\sin (E_1t)\sin (E_2t)\right], \label{L-Rc2}\\
P(\nu_L\to\nu_L^{c})&=&c^4\frac{m_1^2}{E_1^2}\sin^2 (E_1t)+s^4\frac{m_2^2}{E_2^2}\sin^2 (E_2t)
+2c^2s^2\cos (2\phi)\frac{m_1m_2}{E_1E_2}\sin (E_1t)\sin (E_2t), \label{L-Lc}\\
P(\nu_L\to\nu_R)&=&c^2s^2\frac{m_1^2}{E_1^2}\sin^2 (E_1t)+c^2s^2\frac{m_2^2}{E_2^2}\sin^2 (E_2t)
-2c^2s^2\cos(2\phi)\frac{m_1m_2}{E_1E_2}\sin (E_1t)\sin (E_2t). \label{L-R}
\end{eqnarray}
\end{widetext}
We can see that even one generation neutrino oscillations have a very rich structure in the case with
both the Dirac and the Majorana
mass terms and there are various oscillation channels.

$P(\nu_L\to\nu_L)$ expressed by (\ref{L-L1}) and (\ref{L-L2}) is survival probability.
The first line (\ref{L-L1}) is a well-known form and is obtained from also non-relativistic equation.
On the other hand, the second line (\ref{L-L2}) is the correction term originated from the relativistic equation.

$P(\nu_L\to\nu_R^c)$ given by (\ref{L-Rc1}) and (\ref{L-Rc2}) is the probability from active neutrinos $\nu_L$
to sterile neutrinos $\nu_R^c$.
This oscillation mode appears only when both the Dirac and the Majorana mass terms exist.
The probability is the same form as that between two generation neutrinos with different flavors.
Thus, it looks as if the new second-generation neutrino has appeared although we consider one generation.
The term (\ref{L-Rc1}) has the same structure as the transition probability derived from the non-relativistic equation.
On the other hand, the term (\ref{L-Rc2}) newly appeared stands for the relativistic correction.
This oscillation mode is not suppressed by the factor $m^2/E^2$ because of the oscillation without chirality-flip
and may have a large value.
The mixing angle $\theta$ is expressed by bare mass,
\begin{eqnarray}
\tan 2\theta=\frac{m}{2(M_R-M_L)}.
\end{eqnarray}
Accordingly, the angle $\theta$ becomes small in the case $m, M_L \ll M_R$.
In other words, we obtain $\theta \to 0$ in the limit $m_2 \to \infty$ and
the oscillation probability approaches to zero.
This means that the heavy sterile neutrinos are decoupled from the light neutrinos when
the Majorana mass is large and it looks the second generation neutrino does not appear.
There is another possibility that the neutrino oscillation does not occur
by separation of two wave packets corresponding to neutrinos with masses $m_1$ and $m_2$
and the coherent state breaks
when $m_2$ becomes larger than a certain amount compared to $m_1$.
Further discussion is needed on this point but we do not consider detail in this paper.

$P(\nu_L\to\nu_L^{c})$ given by (\ref{L-Lc}) is the probability for the oscillation
from active neutrinos to anti-neutrinos $\nu_L^c$.
This is responsible to 0$\nu\beta\beta$ decay.
This oscillation mode appears if there exists the Majorana mass term.
The probability is suppressed by the factor $m^2/E^2$ because the oscillation accompanies the chirality-flip
and so becomes small due to tiny neutrino mass $m$.
The new CP phase $\phi$ contributes to the probability through the cosine term.

$P(\nu_L\to\nu_R)$ given by (\ref{L-R}) is the probability for the oscillation
from active neutrinos to sterile neutrino $\nu_R$.
The probability is also suppressed in the same way as the above channel.
This oscillation mode appears if there exists the Dirac mass term.
The dependence of the new CP phase $\phi$ is similar to the above channel.

\section{Oscillation Probabilities in Two Generations or More}
In this section, we extend the calculation performed in the previous section to two generations or more.
We also investigate the CP dependence of the probabilities.

The Dirac equation for flavor eigenstates in two generations can be also separated into two parts
related to $\nu$ and $\nu^{\prime}$ as in the case of eq.(\ref{dirac-eq2}).
Below, we consider the part related to $\nu$.
One can calculate the oscillation probabilities for the remaining part in the same way.
The part of $\nu$ in two generations corresponding to (\ref{H-one-gene}) is given by
\begin{widetext}
\begin{eqnarray}
i\frac{d}{dt}\left(\begin{array}{l}
\nu_{\alpha L}^{c} \\ \nu_{\beta L}^{c} \\ \nu_{\alpha R} \\ \nu_{\beta R} \\
\nu_{\alpha L} \\ \nu_{\beta L} \\ \nu_{\alpha R}^{c} \\ \nu_{\beta R}^{c}
\end{array}\right)
=\underbrace{\left(\begin{array}{cccc|cccc}
-p & 0 & 0 & 0 & M_{\alpha\alpha}^L & M_{\alpha\beta}^L & m_{\alpha\alpha} & m_{\beta\alpha} \\
0 & -p & 0 & 0 & M_{\beta\alpha}^L & M_{\beta\beta}^L & m_{\alpha\beta} & m_{\beta\beta} \\
0 & 0 & -p & 0 & m_{\alpha\alpha} & m_{\alpha\beta} & M_{\alpha\alpha}^R & M_{\alpha\beta}^R \\
0 & 0 & 0 & -p & m_{\beta\alpha} & m_{\beta\beta} & M_{\beta\alpha}^R & M_{\beta\beta}^R \\
\hline
M_{\alpha\alpha}^{L*} & M_{\beta\alpha}^{L*} & m_{\alpha\alpha}^* & m_{\beta\alpha}^* & p & 0 & 0 & 0 \\
M_{\alpha\beta}^{L*} & M_{\beta\beta}^{L*} & m_{\alpha\beta}^* & m_{\beta\beta}^* & 0 & p & 0 & 0 \\
m_{\alpha\alpha}^* & m_{\alpha\beta}^* & M_{\alpha\alpha}^{R*} & M_{\beta\alpha}^{R*} & 0 & 0 & p & 0 \\
m_{\beta\alpha}^* & m_{\beta\beta}^* & M_{\alpha\beta}^{R*} & M_{\beta\beta}^{R*} & 0 & 0 & 0 & p
\end{array}\right)}_{H_w}\left(\begin{array}{l}
\nu_{\alpha L}^{c} \\ \nu_{\beta L}^{c} \\ \nu_{\alpha R} \\ \nu_{\beta R} \\
\nu_{\alpha L} \\ \nu_{\beta L} \\ \nu_{\alpha R}^{c} \\ \nu_{\beta R}^{c}
\end{array}\right).
\end{eqnarray}
Let us diagonalize the Hamiltonian $H_w$ and calculate the oscillation probabilities.
As in the case of one generation, we diagonalize the submatrix of the mass part first and next to the whole.
The chirality-flavor eigenstates are obtained by multiplying a unitary matrix $U$ to
the chirality-mass eigenstates as
\begin{eqnarray}
\left(\begin{array}{l}
\nu_{\alpha L}^{c} \\ \nu_{\beta L}^{c} \\ \nu_{\alpha R} \\ \nu_{\beta R} \\
\nu_{\alpha L} \\ \nu_{\beta L} \\ \nu_{\alpha R}^{c} \\ \nu_{\beta R}^{c}
\end{array}\right)
=\underbrace{\left(\begin{array}{cccc|cccc}
U_{\alpha 1}^* & U_{\alpha 2}^* & U_{\alpha 3}^* & U_{\alpha 4}^* & 0 & 0 & 0 & 0 \\
U_{\beta 1}^* & U_{\beta 2}^* & U_{\beta 3}^* & U_{\beta 4}^* & 0 & 0 & 0 & 0 \\
U_{\bar{\alpha} 1}^* & U_{\bar{\alpha} 2}^* & U_{\bar{\alpha} 3}^* & U_{\bar{\alpha} 4}^* & 0 & 0 & 0 & 0 \\
U_{\bar{\beta} 1}^* & U_{\bar{\beta} 2}^* & U_{\bar{\beta} 3}^* & U_{\bar{\beta} 4}^* & 0 & 0 & 0 & 0 \\
\hline
0 & 0 & 0 & 0 & U_{\alpha 1} & U_{\alpha 2} & U_{\alpha 3} & U_{\alpha 4} \\
0 & 0 & 0 & 0 & U_{\beta 1} & U_{\beta 2} & U_{\beta 3} & U_{\beta 4} \\
0 & 0 & 0 & 0 & U_{\bar{\alpha} 1} & U_{\bar{\alpha} 2} & U_{\bar{\alpha} 3} & U_{\bar{\alpha} 4} \\
0 & 0 & 0 & 0 & U_{\bar{\beta} 1} & U_{\bar{\beta} 2} & U_{\bar{\beta} 3} & U_{\bar{\beta} 4}
\end{array}\right)}_{U}
\left(\begin{array}{l}
\nu_{1L}^{c} \\ \nu_{2L}^{c} \\ \nu_{3R} \\ \nu_{4R} \\
\nu_{1L} \\ \nu_{2L} \\ \nu_{3R}^{c} \\ \nu_{4R}^{c}
\end{array}\right).\label{CF-CM2}
\end{eqnarray}
The submatrix related to mass is diagonalized as
\begin{eqnarray}
H_m=U^{\dagger}H_wU =
\left(\begin{array}{cccc|cccc}
-p & 0 & 0 & 0 & m_1 & 0 & 0 & 0 \\
0 & -p & 0 & 0 & 0 & m_2 & 0 & 0 \\
0 & 0 & -p & 0 & 0 & 0 & m_3 & 0 \\
0 & 0 & 0 & -p & 0 & 0 & 0 & m_4 \\
\hline
m_1 & 0 & 0 & 0 & p & 0 & 0 & 0 \\
0 & m_2 & 0 & 0 & 0 & p & 0 & 0 \\
0 & 0 & m_3 & 0 & 0 & 0 & p & 0 \\
0 & 0 & 0 & m_4 & 0 & 0 & 0 & p
\end{array}\right).
\end{eqnarray}
Then, the evolution equation for the chirality-mass eigenstates becomes
\begin{eqnarray}
i\frac{d}{dt}\left(\begin{array}{c}
\nu_{1L}^{c} \\ \nu_{2L}^{c} \\ \nu_{3R} \\ \nu_{4R} \\
\nu_{1L} \\ \nu_{2L} \\ \nu_{3R}^{c} \\ \nu_{4R}^{c}
\end{array}\right)
=\left(\begin{array}{cccc|cccc}
-p & 0 & 0 & 0 & m_1 & 0 & 0 & 0 \\
0 & -p & 0 & 0 & 0 & m_2 & 0 & 0 \\
0 & 0 & -p & 0 & 0 & 0 & m_3 & 0 \\
0 & 0 & 0 & -p & 0 & 0 & 0 & m_4 \\
\hline
m_1 & 0 & 0 & 0 & p & 0 & 0 & 0 \\
0 & m_2 & 0 & 0 & 0 & p & 0 & 0 \\
0 & 0 & m_3 & 0 & 0 & 0 & p & 0 \\
0 & 0 & 0 & m_4 & 0 & 0 & 0 & p
\end{array}\right)\left(\begin{array}{c}
\nu_{1L}^{c} \\ \nu_{2L}^{c} \\ \nu_{3R} \\ \nu_{4R} \\
\nu_{1L} \\ \nu_{2L} \\ \nu_{3R}^{c} \\ \nu_{4R}^{c}
\end{array}\right).
\end{eqnarray}
Furthermore, exchanging some rows and some columns and dividing by each mass
the above equation can be rewritten as
\begin{eqnarray}
i\frac{d}{dt}\left(\begin{array}{c}
\nu_{1L}^{c} \\ \nu_{1L} \\ \nu_{2L}^{c} \\ \nu_{2L} \\
\nu_{3R} \\ \nu_{3R}^{c} \\ \nu_{4R} \\ \nu_{4R}^{c}
\end{array}\right)
=\left(\begin{array}{cc|cc|cc|cc}
-p & m_1 & 0 & 0 & 0 & 0 & 0 & 0 \\
m_1 & p & 0 & 0 & 0 & 0 & 0 & 0 \\
\hline
0 & 0 & -p & m_2 & 0 & 0 & 0 & 0 \\
0 & 0 & m_2 & p & 0 & 0 & 0 & 0 \\
\hline
0 & 0 & 0 & 0 & -p & m_3 & 0 & 0 \\
0 & 0 & 0 & 0 & m_3 & p & 0 & 0 \\
\hline
0 & 0 & 0 & 0 & 0 & 0 & -p & m_4 \\
0 & 0 & 0 & 0 & 0 & 0 & m_4 & p
\end{array}\right)\left(\begin{array}{c}
\nu_{1L}^{c} \\ \nu_{1L} \\ \nu_{2L}^{c} \\ \nu_{2L} \\
\nu_{3R} \\ \nu_{3R}^{c} \\ \nu_{4R} \\ \nu_{4R}^{c}
\end{array}\right).
\end{eqnarray}
If we define
\begin{eqnarray}
C_j=\sqrt{\frac{E_j+p}{2E_j}}, \qquad S_j=\sqrt{\frac{E_j-p}{2E_j}},
\end{eqnarray}
each $2\times 2$ part belong to the same mass is diagonalized as
\begin{eqnarray}
\left(\begin{array}{cc}
C_j & -S_j \\
S_j & C_j
\end{array}\right)
\left(\begin{array}{cc}
-p & m_j \\
m_j & p
\end{array}\right)
\left(\begin{array}{cc}
C_j & S_j \\
-S_j & C_j
\end{array}\right)
=\left(\begin{array}{cc}
-E_j & 0 \\
0 & E_j
\end{array}\right),
\end{eqnarray}
where
\begin{eqnarray}
E_j=\sqrt{p^2+m_j^2} \quad (j=1,2,3,4).
\end{eqnarray}
Accordingly, the chirality-mass eigenstates are related to the energy-helicity eigenstates as
\begin{eqnarray}
\left(\begin{array}{c}
\nu_{1L}^{c} \\ \nu_{1L} \\ \nu_{2L}^{c} \\ \nu_{2L} \\
\nu_{3R} \\ \nu_{3R}^{c} \\ \nu_{4R} \\ \nu_{4R}^{c}
\end{array}\right)
=\underbrace{\left(\begin{array}{cc|cc|cc|cc}
C_1 & S_1 & 0 & 0 & 0 & 0 & 0 & 0 \\
-S_1 & C_1 & 0 & 0 & 0 & 0 & 0 & 0 \\
\hline
0 & 0 & C_2 & S_2 & 0 & 0 & 0 & 0 \\
0 & 0 & -S_2 & C_2 & 0 & 0 & 0 & 0 \\
\hline
0 & 0 & 0 & 0 & C_3 & S_3 & 0 & 0 \\
0 & 0 & 0 & 0 & -S_3 & C_3 & 0 & 0 \\
\hline
0 & 0 & 0 & 0 & 0 & 0 & C_4 & S_4 \\
0 & 0 & 0 & 0 & 0 & 0 & -S_4 & C_4
\end{array}\right)}_{W}
\left(\begin{array}{l}
\nu_{1}^{c-} \\ \nu_{1}^+ \\ \nu_{2}^{c-} \\ \nu_{2}^+ \\
\nu_{3}^- \\ \nu_{3}^{c+} \\ \nu_{4}^- \\ \nu_{4}^{c+}
\end{array}\right). \label{CM-EH2}
\end{eqnarray}
Connecting (\ref{CF-CM2}) and (\ref{CM-EH2}), the chirality-flavor eigenstates are related to
the energy-helicity eigenstates as
\begin{eqnarray}
\left(\begin{array}{l}
\nu_{\alpha L}^{c} \\ \nu_{\beta L}^{c} \\ \nu_{\alpha R} \\ \nu_{\beta R} \\
\nu_{\alpha L} \\ \nu_{\beta L} \\ \nu_{\alpha R}^{c} \\ \nu_{\beta R}^{c}
\end{array}\right)
&=&U
\left(\begin{array}{cccccccc}
1 & 0 & 0 & 0 & 0 & 0 & 0 & 0 \\
0 & 0 & 1 & 0 & 0 & 0 & 0 & 0 \\
0 & 0 & 0 & 0 & 1 & 0 & 0 & 0 \\
0 & 0 & 0 & 0 & 0 & 0 & 1 & 0 \\
0 & 1 & 0 & 0 & 0 & 0 & 0 & 0 \\
0 & 0 & 0 & 1 & 0 & 0 & 0 & 0 \\
0 & 0 & 0 & 0 & 0 & 1 & 0 & 0 \\
0 & 0 & 0 & 0 & 0 & 0 & 0 & 1
\end{array}\right)
W\left(\begin{array}{l}
\nu_{1}^{c-} \\ \nu_{1}^+ \\ \nu_{2}^{c-} \\ \nu_{2}^+ \\
\nu_{3}^- \\ \nu_{3}^{c+} \\ \nu_{4}^- \\ \nu_{4}^{c+}
\end{array}\right) \\
&=&\left(\begin{array}{cccc|cccc}
C_1U_{\alpha 1}^* & S_1U_{\alpha 1}^*
& C_2U_{\alpha 2}^* & S_2U_{\alpha 2}^*
& C_3U_{\alpha 3}^* & S_3U_{\alpha 3}^*
& C_4U_{\alpha 4}^* & S_4U_{\alpha 4}^* \\
C_1U_{\beta 1}^* & S_1U_{\beta 1}^*
& C_2U_{\beta 2}^* & S_2U_{\beta 2}^*
& C_3U_{\beta 3}^* & S_3U_{\beta 3}^*
& C_4U_{\beta 4}^* & S_4U_{\beta 4}^* \\
C_1U_{\bar{\beta}1}^* & S_1U_{\bar{\beta}1}^*
& C_2U_{\bar{\beta}2}^* & S_2U_{\bar{\beta}2}^*
& C_3U_{\bar{\beta}3}^* & S_3U_{\bar{\beta}3}^*
& C_4U_{\bar{\beta}4}^* & S_4U_{\bar{\beta}4}^* \\
C_1U_{\bar{\beta}1}^* & S_1U_{\bar{\beta}1}^*
& C_2U_{\bar{\beta}2}^* & S_2U_{\bar{\beta}2}^*
& C_3U_{\bar{\beta}3}^* & S_3U_{\bar{\beta}3}^*
& C_4U_{\bar{\beta}4}^* & S_4U_{\bar{\beta}4}^* \\
\hline
-S_1U_{\alpha 1} & C_1U_{\alpha 1}
& -S_2U_{\alpha 2} & C_2U_{\alpha 2}
& -S_3U_{\alpha 3} & C_3U_{\alpha 3}
& -S_4U_{\alpha 4} & C_4U_{\alpha 4} \\
-S_1U_{\beta 1} & C_1U_{\beta 1}
& -S_2U_{\beta 2} & C_2U_{\beta 2}
& -S_3U_{\beta 3} & C_3U_{\beta 3}
& -S_4U_{\beta 4} & C_4U_{\beta 4} \\
-S_1U_{\bar{\beta}1} & C_1U_{\bar{\beta}1}
& -S_2U_{\bar{\beta}2} & C_2U_{\bar{\beta}2}
& -S_3U_{\bar{\beta}3} & C_3U_{\bar{\beta}3}
& -S_4U_{\bar{\beta}4} & C_4U_{\bar{\beta}4} \\
-S_1U_{\bar{\beta}1} & C_1U_{\bar{\beta}1}
& -S_2U_{\bar{\beta}2} & C_2U_{\bar{\beta}2}
& -S_3U_{\bar{\beta}3} & C_3U_{\bar{\beta}3}
& -S_4U_{\bar{\beta}4} & C_4U_{\bar{\beta}4}
\end{array}\right)\hspace{-0.cm}
\left(\begin{array}{l}
\nu_{1}^{c-} \\ \nu_{1}^+ \\ \nu_{2}^{c-} \\ \nu_{2}^+ \\
\nu_{3}^- \\ \nu_{3}^{c+} \\ \nu_{4}^- \\ \nu_{4}^{c+}
\end{array}\right).
\end{eqnarray}
Rewriting these relations for fields to those for one particle states, the chirality-mass eigenstates after the
time $t$ are given by
\begin{eqnarray}
&&\hspace{-1cm}|\nu_{\alpha L}^{c}(t)\rangle
=\sum_{j=1,2}\left(C_j U_{\alpha j}e^{iE_jt}|\nu_j^{c-}\rangle
+S_jU_{\alpha j}e^{-iE_jt}|\nu_j^{+}\rangle \right)
+\sum_{j=3,4}\left(C_jU_{\alpha j}e^{iE_jt}|\nu_j^{-}\rangle
+S_jU_{\alpha j}e^{-iE_jt}|\nu_j^{c+}\rangle \right), \\
&&\hspace{-1cm}|\nu_{\alpha R}(t)\rangle
=\sum_{j=1,2}\left(C_j U_{\bar{\alpha} j}e^{iE_jt}|\nu_j^{c-}\rangle
+S_jU_{\bar{\alpha} j}e^{-iE_jt}|\nu_j^{+}\rangle \right)
+\sum_{j=3,4}\left(C_jU_{\bar{\alpha} j}e^{iE_jt}|\nu_j^{-}\rangle
+S_jU_{\bar{\alpha} j}e^{-iE_jt}|\nu_j^{c+}\rangle \right), \\
&&\hspace{-1cm}|\nu_{\alpha L}(t)\rangle
=\sum_{j=1,2}\left(-S_j U_{\alpha j}^*e^{iE_jt}|\nu_j^{c-}\rangle
+C_jU_{\alpha j}^*e^{-iE_jt}|\nu_j^{+}\rangle \right)
+\sum_{j=3,4}\left(-S_jU_{\alpha j}^*e^{iE_jt}|\nu_j^{-}\rangle
+C_jU_{\alpha j}^*e^{-iE_jt}|\nu_j^{c+}\rangle \right), \\
&&\hspace{-1cm}|\nu_{\alpha R}^{c}(t)\rangle
=\sum_{j=1,2}\left(-S_j U_{\bar{\alpha} j}^*e^{iE_jt}|\nu_j^{c-}\rangle
+C_jU_{\bar{\alpha} j}^*e^{-iE_jt}|\nu_j^{+}\rangle \right)
+\sum_{j=3,4}\left(-S_jU_{\bar{\alpha} j}^*e^{iE_jt}|\nu_j^{-}\rangle
+C_jU_{\bar{\alpha} j}^*e^{-iE_jt}|\nu_j^{c+}\rangle \right).
\end{eqnarray}
These can be extended to the case for $n$ generations by replacing
the sum for $j=1,2$ to $j=1,2,\cdots,n$ and the sum for $j=3,4$ to $j=n+1,n+2, \cdots, 2n$.

The oscillation amplitudes are calculated as
\begin{eqnarray}
A(\nu_{\alpha L}\to\nu_{\beta L})&=&\langle \nu_{\beta L}|\nu_{\alpha L}(t)\rangle
=\sum_j U_{\alpha j}^*U_{\beta j}\left(C_j^2e^{-iE_jt}
+S_j^2e^{iE_jt}\right) \\
&=&\sum_j U_{\alpha j}^*U_{\beta j}\left(\frac{E_j+p}{2E_j}e^{-iE_jt}
+\frac{E_j-p}{2E_j}e^{iE_jt}\right)
=\sum_j U_{\alpha j}^*U_{\beta j}\left\{\cos(E_jt)-i\frac{p}{E_j}\sin (E_jt)\right\}, \\
A(\nu_{\alpha L}\to\nu_{\beta R}^c)&=&\langle \nu_{\beta R}^c|\nu_{\alpha L}(t)\rangle
=\sum_j U_{\alpha j}^*U_{\bar{\beta} j}\left(C_j^2e^{-iE_jt}
+S_j^2e^{iE_jt}\right) \\
&=&\sum_j U_{\alpha j}^*U_{\bar{\beta} j}\left(\frac{E_j+p}{2E_j}e^{-iE_jt}
+\frac{E_j-p}{2E_j}e^{iE_jt}\right)
=\sum_j U_{\alpha j}^*U_{\bar{\beta} j}\left\{\cos(E_jt)-i\frac{p}{E_j}\sin (E_jt)\right\}, \\
A(\nu_{\alpha L}\to\nu_{\beta R})&=&\langle \nu_{\beta R}|\nu_{\alpha L}(t)\rangle
=\sum_j U_{\alpha j}^*U_{\bar{\beta} j}^* S_jC_j\left(e^{-iE_jt}-e^{iE_jt}\right) \\
&=&\sum_j U_{\alpha j}^*U_{\bar{\beta} j}^*\frac{m_j}{E_j}(e^{-iE_jt}-e^{iE_jt})
=-i\sum_j U_{\alpha j}^*U_{\bar{\beta} j}^*\frac{m_j}{E_j}\sin (E_jt), \\
A(\nu_{\alpha L}\to\nu_{\beta L}^c)&=&\langle \nu_{\beta L}^c|\nu_{\alpha L}(t)\rangle
=\sum_j U_{\alpha j}^*U_{\beta j}^* S_jC_j\left(e^{-iE_jt}-e^{iE_jt}\right) \\
&=&\sum_j U_{\alpha j}^*U_{\beta j}^*\frac{m_j}{E_j}(e^{-iE_jt}-e^{iE_jt})
=-i\sum_j U_{\alpha j}^*U_{\beta j}^*\frac{m_j}{E_j}\sin (E_jt).
\end{eqnarray}
Furthermore, the oscillation probabilities are obtained by squaring the above amplitudes,
\begin{eqnarray}
P(\nu_{\alpha L}\to\nu_{\alpha L})&=&
1-\sum_j |U_{\alpha j}|^4\cdot \frac{m_j^2}{E_j^2}\sin^2 (E_jt) \nonumber \\
&&\hspace{-1.5cm}-2\sum_{j<k}\left|U_{\alpha j}U_{\alpha k}\right|^2
\left\{2\sin^2 \left(\frac{\Delta E_{jk}t}{2}\right)+\frac{E_jE_k-p^2}{E_jE_k}\sin (E_jt)\sin (E_kt)\right\}, \\
P(\nu_{\alpha L}\to\nu_{\beta L})&=&
-\sum_j |U_{\alpha j}U_{\beta j}|^2\left\{\frac{m_j^2}{E_j^2}\sin^2 (E_jt)\right\} \nonumber \\
&&\hspace{-1.5cm}-2\sum_{j<k}{\rm Re}\left[U_{\alpha j}U_{\beta j}^*U_{\alpha k}^*U_{\beta k}\right]
\left\{2\sin^2 \left(\frac{\Delta E_{jk}t}{2}\right)+\frac{E_jE_k-p^2}{E_jE_k}\sin (E_jt)\sin (E_kt)\right\} \nonumber \\
&&\hspace{-1.5cm}-2\sum_{j<k}{\rm Im}\left[U_{\alpha j}U_{\beta j}^*U_{\alpha k}^*U_{\beta k}\right]
\left\{\sin (\Delta E_{jk}t)-\frac{E_k-p}{E_k}\cos (E_jt)\sin (E_kt)+\frac{E_j-p}{E_j}\cos (E_kt)\sin (E_jt)\right\}, \\
P(\nu_{\alpha L}\to\nu_{\alpha R}^{c})&=&
-\sum_j |U_{\alpha j}U_{\bar{\alpha}j}|^2\left\{\frac{m_j^2}{E_j^2}\sin^2 (E_jt)\right\} \nonumber \\
&&\hspace{-1.5cm}-2\sum_{j<k}{\rm Re}\left[U_{\alpha j}U_{\bar{\alpha}j}^*U_{\alpha k}^*U_{\bar{\alpha}k}\right]
\left\{2\sin^2 \left(\frac{\Delta E_{jk}t}{2}\right)+\frac{E_jE_k-p^2}{E_jE_k}\sin (E_jt)\sin (E_kt)\right\} \nonumber \\
&&\hspace{-1.5cm}-2\sum_{j<k}{\rm Im}\left[U_{\alpha j}U_{\bar{\alpha}j}^*U_{\alpha k}^*U_{\bar{\alpha}k}\right]
\left\{\sin (\Delta E_{jk}t)-\frac{E_k-p}{E_k}\cos (E_jt)\sin (E_kt)+\frac{E_j-p}{E_j}\cos (E_kt)\sin (E_jt)\right\}, \\
P(\nu_{\alpha L}\to\nu_{\beta R}^{c})&=&
-\sum_j |U_{\alpha j}U_{\bar{\beta}j}|^2\left\{\frac{m_j^2}{E_j^2}\sin^2 (E_jt)\right\} \nonumber \\
&&\hspace{-1.5cm}-2\sum_{j<k}{\rm Re}\left[U_{\alpha j}U_{\bar{\beta}j}^*U_{\alpha k}^*U_{\bar{\beta}k}\right]
\left\{2\sin^2 \left(\frac{\Delta E_{jk}t}{2}\right)+\frac{E_jE_k-p^2}{E_jE_k}\sin (E_jt)\sin (E_kt)\right\} \nonumber \\
&&\hspace{-1.5cm}-2\sum_{j<k}{\rm Im}\left[U_{\alpha j}U_{\bar{\beta}j}^*U_{\alpha k}^*U_{\bar{\beta}k}\right]
\left\{\sin (\Delta E_{jk}t)-\frac{E_k-p}{E_k}\cos (E_jt)\sin (E_kt)+\frac{E_j-p}{E_j}\cos (E_kt)\sin (E_jt)\right\},
\end{eqnarray}
\begin{eqnarray}
P(\nu_{\alpha L}\to\nu_{\alpha R})&=&\sum_j \left|U_{\alpha j}U_{\bar{\alpha}j}\right|^2\frac{m_j^2}{E_j^2}\sin^2 (E_jt),
+2\sum_{j<k}{\rm Re}\left[U_{\alpha j}U_{\bar{\alpha}j}U_{\alpha k}^*U_{\bar{\alpha}k}^*\right]\frac{m_jm_k}{E_jE_k}\sin (E_jt)\sin (E_kt), \\
P(\nu_{\alpha L}\to\nu_{\beta R})&=&\sum_j \left|U_{\alpha j}U_{\bar{\beta}j}\right|^2\frac{m_j^2}{E_j^2}\sin^2 (E_jt)
+2\sum_{j<k}{\rm Re}\left[U_{\alpha j}U_{\bar{\beta}j}U_{\alpha k}^*U_{\bar{\beta}k}^*\right]\frac{m_jm_k}{E_jE_k}\sin (E_jt)\sin (E_kt), \\
P(\nu_{\alpha L}\to\nu_{\alpha L}^{c})&=&\sum_j \left|U_{\alpha j}\right|^4\frac{m_j^2}{E_j^2}\sin^2 (E_jt)
+2\sum_{j<k}{\rm Re}\left[U_{\alpha j}^2U_{\alpha k}^{*2}\right]\frac{m_jm_k}{E_jE_k}\sin (E_jt)\sin (E_kt), \\
P(\nu_{\alpha L}\to\nu_{\beta L}^{c})&=&\sum_j \left|U_{\alpha j}U_{\beta j}\right|^2\frac{m_j^2}{E_j^2}\sin^2 (E_jt)
+2\sum_{j<k}{\rm Re}\left[U_{\alpha j}U_{\beta j}U_{\alpha k}^*U_{\beta k}^*\right]\frac{m_jm_k}{E_jE_k}\sin (E_jt)\sin (E_kt),
\end{eqnarray}
\end{widetext}
where $\Delta E_{jk}\equiv E_j-E_k$.
In two generations or more, we can see that the imaginary part of the product of four matrix elements $U$ appears
in the probabilities without chirality-flip.
There are 16 parameters in the $4\times 4$ unitary matrix and the overall phases can be extracted as
\begin{eqnarray}
U\!=\!\left(\!\begin{array}{cccc}
e^{i\rho_{\alpha}} & 0 & 0 & 0 \\
0 & e^{i\rho_{\beta}} & 0 & 0 \\
0 & 0 & e^{i\rho_{\bar{\alpha}}} & 0 \\
0 & 0 & 0 & e^{i\rho_{\bar{\beta}}}
\end{array}\!\!\right)
\!\tilde{U}\!
\left(\!\begin{array}{cccc}
1 & 0 & 0 & 0 \\
0 & e^{i\phi_2} & 0 & 0 \\
0 & 0 & e^{i\phi_3} & 0 \\
0 & 0 & 0 & e^{i\phi_4}
\end{array}\!\!\right),
\end{eqnarray}
where we chose $\phi_1=0$ without loss of generality.
$\tilde{U}$ is the matrix excluding the overall phases and is expressed by $16-7=9$ parameters.
Six parameters are mixing angles and the remaining three are the Dirac CP phases.
The product of four matrix elements that appeared in the oscillation probabilities becomes
\begin{eqnarray}
&&U_{\alpha j}U_{\beta j}^*U_{\alpha k}^*U_{\beta k} \nonumber \\
&&=\tilde{U}_{\alpha j}\tilde{U}_{\beta j}^*\tilde{U}_{\alpha k}^*\tilde{U}_{\beta k}
e^{i(\rho_{\alpha}+\phi_j-\rho_{\beta}-\phi_j-\rho_{\alpha}-\phi_k+\rho_{\beta}+\phi_k)} \nonumber \\
&&=\tilde{U}_{\alpha j}\tilde{U}_{\beta j}^*\tilde{U}_{\alpha k}^*\tilde{U}_{\beta k}.
\end{eqnarray}
In the same way, we also obtain other products,
\begin{eqnarray}
U_{\alpha j}U_{\bar{\alpha} j}^*U_{\alpha k}^*U_{\bar{\alpha} k}
&=&\tilde{U}_{\alpha j}\tilde{U}_{\bar{\alpha} j}^*\tilde{U}_{\alpha k}^*\tilde{U}_{\bar{\alpha} k}, \\
U_{\alpha j}U_{\bar{\beta} j}^*U_{\alpha k}^*U_{\bar{\beta} k}
&=&\tilde{U}_{\alpha j}\tilde{U}_{\bar{\beta} j}^*\tilde{U}_{\alpha k}^*\tilde{U}_{\bar{\beta} k}.
\end{eqnarray}
Thus, the new phases cancel and only the phases included in $\tilde{U}$ appear
in the oscillations without chirality-flip like $\nu_{\alpha L}$ to $\nu_{\beta L}$.
On the other hand, the product included in the oscillation probabilities with chirality-flip
is given by
\begin{eqnarray}
&&U_{\alpha j}U_{\beta j}U_{\alpha k}^*U_{\beta k}^* \nonumber \\
&&=\tilde{U}_{\alpha j}\tilde{U}_{\beta j}\tilde{U}_{\alpha k}^*\tilde{U}_{\beta k}^*
e^{i(\rho_{\alpha}+\phi_j+\rho_{\beta}+\phi_j-\rho_{\alpha}-\phi_k-\rho_{\beta}-\phi_k)} \nonumber \\
&&=\tilde{U}_{\alpha j}\tilde{U}_{\beta j}\tilde{U}_{\alpha k}^*\tilde{U}_{\beta k}^*
e^{2i(\phi_j-\phi_k)}.
\end{eqnarray}
This leads to
\begin{eqnarray}
&&{\rm Re}[\tilde{U}_{\alpha j}\tilde{U}_{\beta j}\tilde{U}_{\alpha k}^*\tilde{U}_{\beta k}^*
e^{2i(\phi_j-\phi_k)}] \nonumber \\
&&={\rm Re}[\tilde{U}_{\alpha j}\tilde{U}_{\beta j}\tilde{U}_{\alpha k}^*\tilde{U}_{\beta k}^*]\cos 2(\phi_j-\phi_k) \nonumber \\
&&-{\rm Im}[\tilde{U}_{\alpha j}\tilde{U}_{\beta j}\tilde{U}_{\alpha k}^*\tilde{U}_{\beta k}^*]\sin 2(\phi_j-\phi_k)
\end{eqnarray}
In the same way, we obtain similar relations on other products as
\begin{eqnarray}
&&{\rm Re}[U_{\alpha j}U_{\bar{\alpha} j}U_{\alpha k}^*U_{\bar{\alpha} k}^*]
={\rm Re}[\tilde{U}_{\alpha j}\tilde{U}_{\bar{\alpha} j}\tilde{U}_{\alpha k}^*\tilde{U}_{\bar{\alpha} k}^*e^{2i(\phi_j-\phi_k)}]
\nonumber \\
&&={\rm Re}[\tilde{U}_{\alpha j}\tilde{U}_{\bar{\alpha} j}\tilde{U}_{\alpha k}^*\tilde{U}_{\bar{\alpha} k}^*]\cos 2(\phi_j-\phi_k)
\nonumber \\
&&-{\rm Im}[\tilde{U}_{\alpha j}\tilde{U}_{\bar{\alpha} j}\tilde{U}_{\alpha k}^*\tilde{U}_{\bar{\alpha} k}^*]\sin 2(\phi_j-\phi_k) \\
&&{\rm Re}[U_{\alpha j}U_{\bar{\beta} j}U_{\alpha k}^*U_{\bar{\beta} k}^*]
={\rm Re}[\tilde{U}_{\alpha j}\tilde{U}_{\bar{\beta} j}\tilde{U}_{\alpha k}^*\tilde{U}_{\bar{\beta} k}^*e^{2i(\phi_j-\phi_k)}] \nonumber \\
&&={\rm Re}[\tilde{U}_{\alpha j}\tilde{U}_{\bar{\beta} j}\tilde{U}_{\alpha k}^*\tilde{U}_{\bar{\beta} k}^*]\cos 2(\phi_j-\phi_k) \nonumber \\
&&-{\rm Im}[\tilde{U}_{\alpha j}\tilde{U}_{\bar{\beta} j}\tilde{U}_{\alpha k}^*\tilde{U}_{\bar{\beta} k}^*]\sin 2(\phi_j-\phi_k)
\end{eqnarray}
Thus, the probabilities for oscillations with chirality-flip depend on the new CP phases in addition to the Dirac CP phases included in $\tilde{U}$.
In two generations, there are three independent differences between $\phi_j$ and $\phi_k$,
for example $\phi_1-\phi_2$, $\phi_1-\phi_3$ and $\phi_1-\phi_4$.
We can determine the values of $2(\phi_j-\phi_k)$ uniformly from the range of 0 degrees and 360 degrees in principle, 
because both the sine and the cosine terms of $2(\phi_j-\phi_k)$ are included in the probabilities for the oscillations with chirality-flip.

Here, let us count the number of independent phases that appeared in our formulation.
In the case of $n$ generations, $U$ given above becomes $2n\times 2n$ complex unitary matrix.
It has each $4n^2$ real and imaginary components.
Unitary condition gives $\frac{2n(2n-1)}{2}=n(2n-1)$ constraints for imaginary part.
Furthermore, $2n$ overall phases $\rho$ do not appear in the oscillation probabilities.
These phases are corresponding to those absorbed in the redefinition of the charged lepton fields.
The oscillation probabilities depend on the phases $\phi$ in the form of a difference $\phi_i-\phi_j$
and the number of independent parameters included in the probabilities is $2n-1$.
These phases are corresponding to the Majorana phases multiplying to $\tilde{U}$ from outside.
On the other hand, the number of the Dirac phases included in $\tilde{U}$
is calculated as $4n^2-n(2n-1)-2n-(2n-1)=2n^2-3n+1=(2n-1)(n-1)$.
We summarize the cases of $n=1,2,3$ and general $n$ generations in Table I.

\vspace{0.5cm}

\begin{widetext}
\begin{center}
Table I : Independent parameters in $U$ (Case for the flavors of $\nu_R$ to be distinguishable)
\end{center}

\begin{center}
$\begin{array}{|c||c|c|c|c|c|}
\hline
{\rm generation} & {\rm Majorana \,\, CP \,\, phase} & {\rm Dirac \,\, CP \,\, phase} &
{\rm Total \,\, number \,\, of \,\, CP \,\, phase} \\\hline\hline
1 & 1 & 0 & 1 \\\hline
2 & 3 & 3 & 6 \\\hline
3 & 5 & 10 & 15 \\\hline
n & 2n-1 & (2n-1)(n-1) & n(2n-1) \\\hline
\end{array}$ \\

\end{center}
\end{widetext}
\vspace{0.5cm}

Next, let us consider the case that the flavor of the right-handed neutrinos $\nu_R$ and $\nu_R^c$
cannot be distinguished even beyond the Standard Model.
Namely, $\nu_R$ and $\nu_R^c$ are sterile neutrinos in the true sense because they
interact only through gravity.
In this case, individual information of the subscripts $\bar{\alpha}$ and $\bar{\beta}$
cannot be measured and only the sum about these subscripts can be obtained.
If the right-handed neutrino is singlet for weak interactions,
the number of generations is not always the same as that for left-handed neutrino.
Here, we define the number of generations for left-handed neutrino (active neutrino) as $n_L$
and that for right-handed neutrino (sterile neutrino) as $n_R$
and calculate the number of independent CP phases.

In this case, we can distinguish $n_L$ left-handed flavors on the subscript $\alpha$,
and $n_L+n_R$ mass eigenvalues on the subscript $i$.
Namely, $n_L\times (n_L+n_R)$ mixing submatrix can be measured.
There are each $n_L(n_L+n_R)$ real and imaginary components.
The imaginary components are constrained by $\frac{n_L(n_L-1)}{2}$ unitary conditions.
Moreover, $n_L$ phases are absorbed by the redefinition of the charged leptons.
Namely, the number of independent CP phases is calculated as
$n_L(n_L+n_R)-\frac{n_L(n_L-1)}{2}-n_L=\frac{n_L^2+2n_Ln_R-n_L}{2}
=\frac{1}{2}n_L(n_L+2n_R-1)$.
$n_L+n_R-1$ phases of these are out of $\tilde{U}$ and are
corresponding to the Majorana phases.
On the other hand, the number of the Dirac phases included in $\tilde{U}$ is given by
the difference of the number of the total and the Majorana phases as
$\frac{1}{2}n_L(n_L+2n_R-1)-(n_L+n_R-1)=\frac{1}{2}(n_L-1)(n_L+2n_R-2)$.
In addition to the general case, we give some examples in Table II.
The results in Table I can be obtained by taking the number of active neutrinos as $n_L=2n$
and the number of sterile neutrinos as $n_R=0$ in Table II.
The number of independent phases was discussed in previous papers, for example, see Schechter and Valle \cite{Schechter-Valle},
Endoh et.al. \cite{Endoh-Morozumi} and Rodejohann and Valle \cite{Rodejohann-Valle}.
The results obtained in Table I and Table II are consistent with those in the above references.

\vspace{0.5cm}

\begin{widetext}
\begin{center}
Table II : Independent parameters in $U$ (Case for the flavors of $\nu_R$ to be non distinguishable)
\end{center}

\begin{center}
$\begin{array}{|c||c|c|c|c|c|}
\hline
{\rm generation} & {\rm Majorana \,\, CP \,\, phase} & {\rm Dirac \,\, CP \,\, phase} &
{\rm Total \,\, number \,\, of \,\, CP \,\, phase} \\\hline\hline
1+1 & 1 & 0 & 1 \\\hline
2+2 & 3 & 2 & 5 \\\hline
3+1 & 3 & 3 & 6 \\\hline
3+2 & 4 & 5 & 9 \\\hline
3+3 & 5 & 7 & 12 \\\hline
n_L+n_R & n_L+n_R-1 & \frac{1}{2}(n_L-1)(n_L+2n_R-2) & \frac{1}{2}n_L(n_L+2n_R-1) \\\hline
\end{array}$ \\

\end{center}
\end{widetext}

\vspace{0.5cm}

In Table I and II, we consider the case of non-zero Majorana masses for $\nu_L$.
However, the triplet Higgs giving the Majorana masses to $\nu_L$ has not been observed until now
and it seems likely that the Majorana masses for three active neutrinos are completely zero.
In this case, the number of independent phases is reduced \cite{Schechter-Valle, Endoh-Morozumi}.
If there is no triplet Higgs and there exists only one right-handed neutrino, namely in the $3+1$ model,
two of the mass eigenvalues for light active neutrinos vanish and we cannot explain the experimental results so far.
Furthermore, in the case that there are two right-handed neutrinos, one of the mass eigenvalues vanishes.
Although the experimental results so far may be explained, a strong constraint is imposed.
Thus, if triplet Higgs exists, it must be discovered in near future and if it does not find,
the Majorana masses for left-handed neutrinos are strongly constrained.

\section{Summary}

In this paper, we have investigated the neutrino oscillation probabilities in the case with
both the Dirac and the Majorana masses.
If two kinds of mass terms exist, the left-handed neutrino $\nu_L$ and the right-handed anti-neutrino $\nu_R^c$
are included in the same multiplet and can change each other.
Therefore, we can understand the sterile neutrinos in the framework of three generations naturally.
$\nu_L \to \nu_R^c$ and $\nu_L^c \to \nu_R$ oscillations realized in this formulation
are not suppressed by the factor $m^2/E^2$ because of the oscillations without chirality-flip.
Accordingly, we can explain some experimental results suggesting the existence of sterile neutrinos
naturally.
If the oscillations to sterile neutrinos in vacuum will be confirmed,
we can interpret it as a strong suggestion that neutrinos are the Majorana particle and they have both the Dirac and the Majorana
mass terms.
In many previous papers, sterile neutrinos were introduced independently from the other three active light neutrinos as singlet
in weak interactions and were called the fourth-generation neutrino.
However, it seems more natural
that the sterile neutrinos are not independent of other active neutrinos $\nu_L$ and $\nu_L^c$
but they exist in pairs with active neutrinos.
Namely, we think that the light ones of $\nu_R$ and $\nu_R^c$ included in the three-generation framework
play the role of the fourth, fifth, or sixth-generation neutrinos.


\end{document}